\documentclass[]{spie}  
\usepackage[]{graphicx}

\title{ALMA service data analysis and level 2 quality assurance with CASA} 

\author{Dirk Petry\supit{a}, Baltasar Vila-Vilaro\supit{b}, Eric Villard\supit{b}, Shinya Komugi\supit{c}, and Scott Schnee\supit{d}
\skiplinehalf
\supit{a} European Southern Observatory, Karl-Schwarzschild-Str. 2, 85748 Garching, Germany\\
\supit{b} Joint ALMA Observatory, Alonso de Cordova 3107, Vitacura, Santiago, Chile\\
\supit{c} National Astronomical Observatory of Japan, 2-21-1 Osawa, Mitaka, Tokyo 181-001, Japan\\
\supit{d} National Radio Astronomy Observatory, 520 Edgemont Rd, Charlottesville, VA, U.S.A. 
}

\authorinfo{Further author information: (Send correspondence to D. Petry)\\E-mail: dpetry@eso.org, Telephone: +49 (0)89 3200 6511
}

\begin{document} 
  \maketitle 

\begin{abstract}
The Atacama Large mm and sub-mm Array (ALMA) radio observatory is one of the world's 
largest astronomical projects. After the very successful conclusion of the first observation 
cycles Early Science Cycles 0 and 1, the ALMA project can report many successes and 
lessons learned. The science data taken interleaved with commissioning tests for the still 
continuing addition of new capabilities has already resulted in numerous publications 
in high-profile journals.  The increasing data volume and complexity are challenging but 
under control. The radio-astronomical data analysis package "Common Astronomy Software Applications"
(CASA) has played a crucial role in this effort. This article describes the implementation 
of the ALMA data quality assurance system, in particular the level 2 which is based on CASA, 
and the lessons learned. 
\end{abstract}




\section{INTRODUCTION}
\label{sec:intro}
The Atacama Large Millimeter/submillimeter Array (ALMA)
is a major new astronomical observatory. It started operation in 2011 and
was officially inaugurated in 2013. ALMA consists of an array of fifty 12-m antennas
and an additional compact array (ACA) of twelve 7-m and four 12-m antennas to enhance 
ALMA's ability to image extended targets. Among the 12-m antennas of the interferometric array,
baseline lengths up to 16~km will be achieved. 

The ALMA project is an international collaboration between Europe, East Asia, and 
North America in cooperation with the Republic of Chile.
The official project website for scientists is the {\it ALMA Science Portal}
{\tt http://www.almascience.org}. A more detailed description can be found
in the proceedings of a parallel session at this conference \cite{cox} and references therein. 

ALMA has been performing ``Early Science'' observations with a subset of its antennas
since 2011. The many capabilities are gradually commissioned.
In the first two observing proposal cycles (Cycles 0 and 1), ALMA offered
receivers that covered four separate atmospheric windows 
(bands 3, 6, 7, and 9) 
in the range between 84~GHz to 720~GHz (3~mm to 0.42~mm). In subsequent cycles, 
more receiver bands will become available. 
Cycle 0 observations resulted in 78 refereed publications so far.
Cycle 1 has ended on 31 May 2014, i.e. shortly before the beginning of this conference.
Its science impact cannot fully be measured, yet.

In Cycle 1, typically 30 of the 50 antennas of the 12-m array were used for science observations.  
From the ACA, typically 8 antennas were available. The last of
the 66 antennas was brought up to the observatory site in March 2014 and test observations with
more than 50 antennas have already taken place.
When the full set of antennas is in operation, the average daily science data volume 
generated by ALMA is expected to be roughly one TByte.  
All ALMA data is stored in the {\it ALMA Science Data Model} (ASDM) format 
which in its present implementation is a collection of XML and binary MIME
encoded files.

When a team of scientists (led by the principle investigator, the ``PI'') proposes an
observation with ALMA, they are not applying for a certain amount of observation time
but they are asking for the achievement of a {\it Science Goal}.
This is a sensitivity and observation setup requirement which the observatory pledges to meet
if the proposal is accepted. 

The observations belonging to a certain proposal are regarded to be completed when
the Science Goal(s) defined in the proposal are achieved. In order to confirm that
achievement, a full calibration and at least partial imaging of the science data is necessary. 
This analysis is carried out as part of the ALMA {\it Quality Assurance} (QA). 

The ALMA operations are based on {\it service observing}. 
This means firstly that the scientists who proposed the observations
and who will have the proprietary rights on the data for the first year after they were taken,
are not required (nor permitted) to be present during the observation.
Instead, the observations are carried out using dynamic scheduling, i.e. matching 
the requirements set in the proposal with the present
status of the observatory, the atmospheric conditions, and the local siderial time.

It is furthermore the ALMA project's ambition to make the observatory available to all 
astronomers, not only sub-mm radio astronomy experts.
The observatory therefore provides {\it service data analysis}. As part of the QA work,
personnel at  the Joint ALMA Observatory in Chile (JAO) and the three ALMA Regional Centers (ARCs) 
perform detailed calibration and imaging on all observations. The PI obtains the resulting
calibration and imaging scripts and imaging products \cite{qa2products}.

In the simplest case, if the PI is content with the imaging products, he/she can 
proceed to a publication based on the obtained imaging products
without worrying about learning the use of the ALMA data analysis software.
For more complex observing projects, the PI may need to put more work 
into at least the imaging, sometimes also the calibration. In any case, the standard ALMA imaging 
products and calibration tables will be archived and made available to the public (together with 
the raw data and all analysis scripts) as soon as the PI's proprietary time ends.   

The main software tool in the processing of ALMA data is CASA (``Common Astronomy Software Applications''),
a package which is developed together by all ALMA partners under NRAO management.

This article summarises the design of the ALMA service data analysis 
and QA infrastructure and describes how CASA is employed in this context.
Detailed descriptions of many of the processes mentioned here can be found
in the ALMA Technical Handbook \cite{handbook2} which is updated for every observing cycle.
Also, there are several other related articles to be found in the different 
proceedings volumes of this conference. See in particular Ref.\citenum{schnee2014}.

\section{The ALMA Pipelines}

To achieve high data quality, ALMA employs three data analysis pipelines at different stages
of the observing process: 
\begin{enumerate}
\item The Telescope Calibration pipeline (TelCal)
\item The Quicklook Display pipeline
\item The offline data analysis (including the Science Pipeline, simply called ``The Pipeline'')
\end{enumerate}
The first two of these are running during and immediately after the observations (online). 
They produce measurements of the characteristic parameters of the observatory
for the {\it Observatory State and Calibration Database} and permit
the Astronomer on Duty (AoD) to monitor the data quality. 

Due to the real-time processing constrains and the high data volume, TelCal and Quicklook in many cases
cannot work on the data at full spectral resolution nor perform detailed high-resolution imaging. 
Instead they mostly work on averaged data.
The full performance of ALMA is achieved in the offline analysis which is (trivially) parallelised in several
ways and therefore able to handle the instrument's native temporal, spectral, and spatial resolution.
This offline analysis is based on the CASA software package which is described in the next section.

\section{CASA technical summary}

The offline data analysis needs a software package 
capable of handling all features of the ALMA data including the high spectral resolution and sensitivity
and the large data volume. In 2003, the {\it CASA} package was selected 
by the project as the ALMA data reduction software.
The structure of CASA was already described in Refs. \citenum{mcmullin_2007} and \citenum{petry_2012}. It has not changed significantly since 2012. So only 
a brief summary is given here. See also Ref. \citenum{ott_2014}. 

CASA (Common Astronomy Software Applications) is a general package for all
radio-astronomical data analysis. The development team consists of scientists
and software engineers from NRAO, ESO, and NAOJ. The main focus is on the
support for the ALMA and the Carl Jansky VLA observatories.

The CASA management, the project scientist, and the subsystem scientists from the ALMA
and JVLA side are all based at NRAO. At ESO and NAOJ so-called cognizant leads
accompany the development for the ALMA side.
Documentation is available from the CASA homepage {\tt http://casa.nrao.edu}. 

The lower-level CASA functionality is implemented using C++.
The fundamental libraries of CASA are collected in a sub-package named {\tt casacore}
which extends the C++ Standard Template Library
providing methods for the handling of files, physical and astronomical 
quantities, coordinate systems and reference frames, advanced mathematical
operations, and more. {\tt casacore} is developed in collaboration with ASTRON
and ATNF and is documented on googlecode at {\tt http://code.google.com/p/casacore/}.
The CASA team presently still keeps its own copy of a {\tt casacore} repository
with regular mergers to the googlecode version. It is planned to end this
duplication soon and converge to a single version.

{\tt casacore} with its rich and well-tested set of basic C++ classes is  
useful for any astronomical software application, not only for radio astronomy.

Based on {\tt casacore}, CASA implements in a separate sub-package the functionality
needed specifically for radio astronomy. This includes
calibration algorithms based on the Measurement Equation \cite{hamakeretal_1996},
interferometric imaging algorithms, a powerful viewer tool to visualise imaging results,
image analysis algorithms, and a simulator for radio interferometers.
In addition, a complete set of algorithms for the analysis of single-dish (non-interferometric) 
radio data is provided by including the ASAP package \cite{asap} with special extensions.

In a third sub-package, the user interface and high-level analysis functionality
is implemented by binding the functionality implemented in C++ to the {\it Python} scripting language
and adding a layer of high-level ``tasks'' which are Python scripts implementing common
procedures used in interferometric and single-dish radio data analysis.  

Finally, the employment of the {\it iPython} package creates the command line user
interface which gives CASA the look and feel which users of modern data analysis packages like, e.g., MatLab or IDL
are used to.

The supported computer platforms of CASA in its latest version 4.2.2 are the common Linux 
distributions (64 bit only) and Mac OSX 10.7 and 10.8. The binary release and the source code are 
available under GNU Public License (see {\tt http://casa.nrao.edu}).

\section{ALMA Quality Assurance}

The goal of ALMA Quality Assurance (QA) is to deliver to the PI a
reliable final data product that has reached the desired control
parameters outlined in the science goals and that is calibrated to the
desired accuracy and free of calibration or imaging artifacts.

As mentioned in the introduction, ALMA follows a paradigm of ``Science-goal-oriented service data analysis''.
This means the PI defines science goals in the proposal aided by a specially developed ALMA software 
package, the Observing Tool (OT).
This formalised proposal defines so-called ``Scheduling Blocks'' (SBs).
An SB is the prototype of an atomic (ca. 0.5 h -  1 h) observation to reach a specific science goal.
SBs are stored in a special XML format which can be directly understood by the observation scheduler.
An execution of an SB, i.e. the actual observation, is called an ``Exec Block'' (EB).
By executing one SB several times, many EBs can be produced and hence sensitivity accumulated.

SBs which depend on each other because their data analysis has to be carried out together,
form an ``Observation Unit Set'' (OUS). The way ALMA projects are set up at the moment,
typically every OUS only contains one SB.
The number of required EBs to reach the science goal of the OUS is estimated by the OT in
an analytic calculation by the so-called sensitivity calculator which takes into account detailed
models of the observatory, the atmosphere, and the calibration sources.
It is the task of the QA to verify that each EB was properly executed and that
the OUS (after the nominal number of good EBs is reached) does indeed achieve the science goal.
ALMA QA consists of 3 (+1) steps (see also figure \ref{fig:almadataflow}):

   \begin{figure}
   \begin{center}
   \begin{tabular}{c}
   \includegraphics[width=13.5cm]{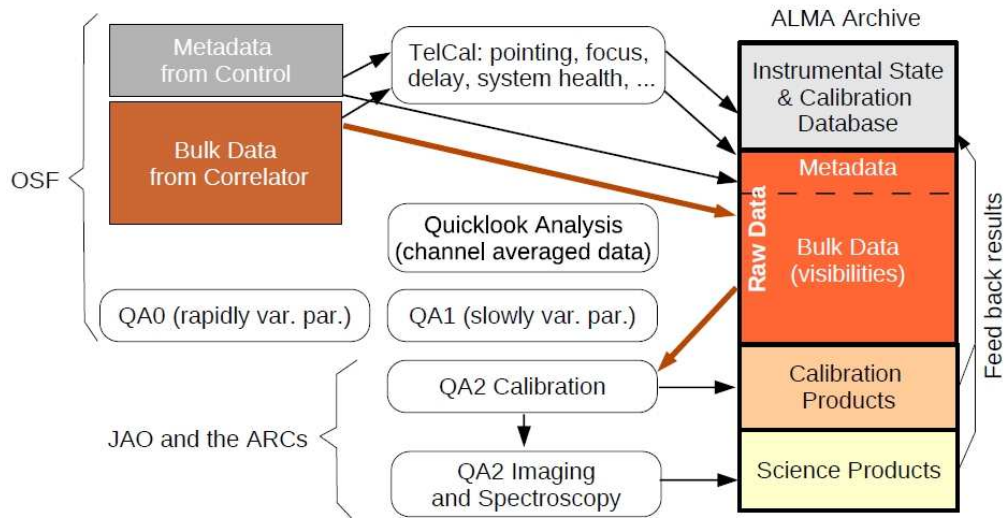}
   \end{tabular}
   \end{center}
   \caption[almadataflow] 
   { \label{fig:almadataflow} 
     Schematic overview of the ALMA data flow indicating the various QA stages.
     For simplicity, the detailed data flow to and from Quicklook, QA0, and QA1
     has been omitted. All these stages access the archive.
   }
   \end{figure} 

\begin{description}
\item[QA0:] Immediate checks of data quality at the time of the observation or shortly after:
            Atmosphere, Antennas, Front-Ends, Connectivity, Back-Ends
\item[QA1:] Monitor slowly varying array performance parameters (Arrays, Antennas, Calibration Sources)
            and also rapid changes which affect the entire array
\item[QA2:] The completion of an OUS triggers QA2 to confirm that the Science Goal was met.
            If not, additional EBs will need to be obtained. QA2 implies full calibration and 
            generation of official science products. After QA2 is complete (Science Goal met), 
            the data is delivered to the PI.
\item[QA3:] If there are problems found with the data after the delivery, the PI or the ALMA contact
            scientist of the PI can file a problem report via the ALMA helpdesk which triggers QA3.
            The problem is examined and a solution searched for. Re-reduction of the data may be performed,
            possibly replacement of products in the archive.
\end{description}

In the following we concentrate on the QA2 step.
For every OUS which has reached the number of planned executions passing QA0 and QA1,
QA2 proceeds to perform a full calibration.
The resulting CASA calibration tables and scripts for the OUS are archived.
Furthermore, imaging is performed (a) to the point where it can be decided whether
each Science Goal was met and (b) to obtain a standard set of science products
for archiving, i.e. for later archival research (see Ref. \citenum{qa2products}).
Ultimately, QA2 on all data from standard observing is supposed to be performed
by the fully automated Pipeline as mentioned above. The Pipeline is still under 
development and a first version of it is ready to be put in service at the beginning 
of Cycle 2. Until the Pipeline can process data from all possible observing modes,
two analysis paths will coexist in QA2:
\begin{enumerate}
\item Semi-automatic processing (calibration and imaging) using the so-called Script Generator 
\item Automated calibration with the Pipeline followed by semi-automatic imaging
\end{enumerate}

\subsection{ALMA QA2 with the Script Generator}

Before a fully automated data analysis pipeline can be commissioned, the manual data analysis 
has to be understood. The necessary expertise to design the data analysis procedures
grows as the operators gain experience with the instrument. On the other hand, as a new 
observatory starts operation, resources for software development are stretched thin and, 
what is more, there is strong pressure to produce results quickly and open the observatory 
to the community for actual science observations.

A naive, linear approach which waits for the completion of the observatory construction
and commissioning before detailed work on data analysis automation is started,
leads to intolerable delays in the development of the analysis pipeline.
Already the task of processing test observations is overwhelming if not aided by sufficient
software tools. On the other hand, if the development is started too early,
the lack of experience with the real system can lead to wasted efforts and the need to
redesign.

The solution is to {\it gradually} automate the data analysis as new observing modes become
available during the completion of the observatory commissioning.
The gradual automation is achieved by having a basic toolkit for analysis 
from which a prototype pipeline can be built for each observing mode.
Once the analysts agree that the prototype meets all requirements, it can be
transferred into the final, fully automated Pipeline. In this way the Pipeline
acquires more and more capabilities and can take over more and more of
the QA2 work.

For ALMA, the role of the basic analysis toolkit is played by CASA with its
scripting language Python.
A large team of ALMA scientists worked together to develop the initial
best practices to perform a robust standard calibration of ALMA Cycle 0 data.
Based on these, the pipeline prototyping could simply have been performed
by developing separate CASA scripts 
for each observing mode. However, the number of possible combinations of
setups was regarded as too large, such that an automated assistant to the analyst
was thought to be necessary.
This additional software tool is called the Script Generator.
It was first devised and implemented by Eric Villard (JAO) and is still 
maintained by him.

   \begin{figure}
   \begin{center}
   \begin{tabular}{c}
   \includegraphics[width=14cm]{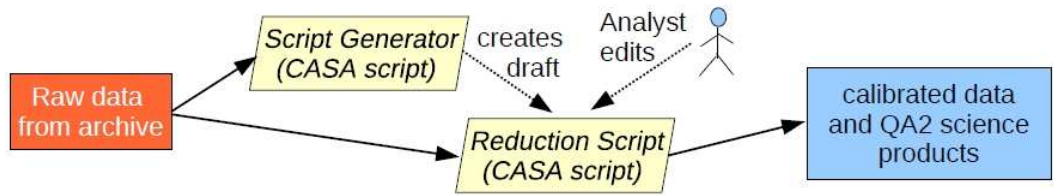}
   \end{tabular}
   \end{center}
   \caption[qa2-w-scriptgen] 
   { \label{fig:qa2-w-scriptgen} 
     Illustration of the script generator principle. With the help of the Script Generator,
     the data analyst generates first a draft analysis script. The Script Generator 
     parses the data to be analysed and implements best practices and previous 
     analysis experience. The analyst then manually finalises the script adapting 
     it (if necessary) to unforeseen features of the data. }
   \end{figure} 

The Script Generator (see figure \ref{fig:qa2-w-scriptgen}) is itself a CASA script.
It takes as only input the dataset to be analysed.
Based on the properties of the dataset, it selects the appropriate analysis path
and creates a new CASA draft analysis script specific for the dataset.
This draft script can in some cases already be final.
In other cases, the analyst will have to make small modifications to the script
until it accommodates all special features of the dataset.
These concern mostly shortcomings of the data caused by hardware problems.
In addition to providing analysis script drafts for the calibration,
the present ALMA Script Generator also provides scripts (a) for combining
several EBs into one final dataset ready for imaging and (b) for the
imaging itself.

The script generator approach has proven to be highly effective.
The analysts save time on repetitive tasks but keep full control as each 
step of the analysis is readily visible and can quickly be modified.
The analysis does not become a black box which is blindly trusted.
Solutions to newly encountered problems and improvements can easily be fed back.
The mature Script Generator helps new team members to learn quickly
while saving time also for very experienced analysts who are enabled
to concentrate on the new and difficult aspects of the data.
As confidence increases, the automation can be increased step by step.

\subsection{ALMA QA2 with the Pipeline}

Also the fully automated ALMA Pipeline is based on CASA and is planned
to eventually become part of it as a set of additional, ALMA-specific
high-level functionality ``tasks''.
It is meant to ultimately process both interferometry and single-dish data
in a fully automated, data-driven way including the imaging.
While it is mainly meant to be run in a high-performance computing environment,
the present design foresees that any CASA user will be able to run it
on a standard, reasonably powerful workstation with sufficient disk space
to hold the raw data (typically up to 1 TB for a simple project) and
the intermediate and final data products (up to 3 TB). 

As described in the previous section, the Pipeline development is fed
by the experience gained in manual and Script-Generator-assisted analysis.
This experience is encoded in so-called ``heuristics''. The Pipeline
(and also the Script Generator) can be regarded as an expert system
for ALMA data analysis.
Pipeline commissioning and verification is performed by comparing the 
results with those obtained from the Script-generator-assisted analysis.

The first actual use of the Pipeline for QA2 will take place in
ALMA Cycle 2 in the second half of 2014. It will be employed to perform
the calibration for datasets from well understood observing modes.
Imaging will initially still be done only with Script Generator assistance
and (like all other analysis) gradually be moved over into the Pipeline.

\subsection{ALMA QA2 statistics in Cycle 1}

At the time of writing, ALMA Cycle 1 is nearly complete.
The QA2 effort was significant.
So far, the (at any time) ca. 40 QA2 analysts at the Joint ALMA Observatory (JAO) and the
three ALMA Regional Centres (ARCs) in East Asia, Europe, and North America
have processed more than 980 h of observations in Cycle 1 since January 2013
and delivered more than 188 Observing Unit Sets to their PIs.
The typical (median) analysis time in Script-Generator-assisted QA2 is 3 to 4 working days
per Execution Block. This includes data transfer and packaging.
Only very few cases of QA3 have occurred so far.
The observatory has received enthusiastic feedback from the PIs concerning data quality. 

\section{SUMMARY}

ALMA is working well but commissioning will still have to continue
interleaved with science observations for a number of cycles until all
initially planned capabilities are available on the full set of antennas.
The service data analysis model has proven to be viable and flexible
enough to cope with the varying conditions.
PIs are excited about the achieved data quality.
The overall QA2 system is running better and better but will remain labour-intensive
until the fully automated Pipeline has implemented heuristics for at least most of the
planned observing modes.
The work spent on QA2, however, is not wasted but creates the necessary
knowledge and expertise on ALMA data analysis in a broad base of
the ALMA personnel.

\bibliography{SPIE-9152-17-dpetry}   
\bibliographystyle{spiebib}

\end{document}